\documentclass[a4paper,reprint,amsfonts,amssymb,amsmath,superscriptaddress,twocolumn]{revtex4-2}
\usepackage{graphicx, hyperref,float}
\hypersetup{colorlinks,citecolor=blue,urlcolor=blue,linkcolor=blue}
\usepackage[all]{hypcap}
\usepackage[version = 4]{mhchem}
\usepackage{lipsum}  
\usepackage{orcidlink}
\usepackage{academicons}
\usepackage{xcolor}
\usepackage{soul}
\setstcolor{red}
\newcommand{\orcid}[1]{\href{https://orcid.org/#1}{\textcolor[HTML]{A6CE39}{\aiOrcid}}}

\begin{document}

% Title
\title{Current-phase relation of a short multi-mode Bi$_2$Se$_3$ topological insulator nanoribbon Josephson junction with ballistic transport modes}

%%%%%%%%%%%%%%%%%% Author list %%%%%%%%%%%%%%%

% Ananthu
\author{Ananthu P. Surendran\orcidlink{0000-0002-0949-4145}}
\affiliation{Quantum Device Physics Laboratory, Department of Microtechnology and Nanoscience, Chalmers University of Technology,
SE-41296 Göteborg, Sweden}

% Domenico
\author{Domenico Montemurro\orcidlink{0000-0001-8944-0640}}
\affiliation{Quantum Device Physics Laboratory, Department of Microtechnology and Nanoscience, Chalmers University of Technology,
SE-41296 Göteborg, Sweden}
\affiliation{ Dipartimento di Fisica “Ettore Pancini,” Università degli Studi di Napoli Federico II, I-80125 Napoli, Italy}

% Gunta
\author{Gunta Kunakova\orcidlink{0000-0003-0243-2678}}
\affiliation{Quantum Device Physics Laboratory, Department of Microtechnology and Nanoscience, Chalmers University of Technology, SE-41296 Göteborg, Sweden}
\affiliation{Institute of Chemical Physics, University of Latvia, Raina Blvd. 19, LV-1586 Riga, Latvia}

% Xavier
\author{Xavier Palermo\orcidlink{0000-0001-9997-3053}}
\affiliation{Quantum Device Physics Laboratory, Department of Microtechnology and Nanoscience, Chalmers University of Technology,
SE-41296 Göteborg, Sweden}

% Kiryl
\author{Kiryl Niherysh\orcidlink{0000-0002-9861-9957}}
\affiliation{Quantum Device Physics Laboratory, Department of Microtechnology and Nanoscience, Chalmers University of Technology,
SE-41296 Göteborg, Sweden}
\affiliation{Institute of Chemical Physics, University of Latvia, Raina Blvd. 19, LV-1586 Riga, Latvia}

% Edo
\author{Edoardo Trabaldo\orcidlink{0000-0002-0188-6814}}
\affiliation{Quantum Device Physics Laboratory, Department of Microtechnology and Nanoscience, Chalmers University of Technology,
SE-41296 Göteborg, Sweden}

% Dima
\author{ Dmitry S. Golubev\orcidlink{0000-0002-0609-8921}}
\affiliation{QTF Centre of Excellence, Department of Applied Physics, Aalto University, P.O. Box 15100, FI-00076 Aalto, Finland}

% Jana
\author{Jana Andzane\orcidlink{0000-0002-9802-6895}}
\affiliation{Institute of Chemical Physics, University of Latvia, Raina Blvd. 19, LV-1586 Riga, Latvia}

% Donats Erts
\author{Donats Erts\orcidlink{0000-0003-0345-8845}}
\affiliation{Institute of Chemical Physics, University of Latvia, Raina Blvd. 19, LV-1586 Riga, Latvia}

% Floriana
\author{Floriana Lombardi\orcidlink{0000-0002-3478-3766}}
\affiliation{Quantum Device Physics Laboratory, Department of Microtechnology and Nanoscience, Chalmers University of Technology,
SE-41296 Göteborg, Sweden}

%Thilo
\author{Thilo Bauch\orcidlink{0000-0002-8918-4293}}
\email[e-mail: ]{thilo.bauch@chalmers.se }
\affiliation{Quantum Device Physics Laboratory, Department of Microtechnology and Nanoscience, Chalmers University of Technology,
SE-41296 Göteborg, Sweden}

\date{\today}

%%%%%%%%%%%%%%%%%%%%%%% Abstract %%%%%%%%%%%%%%%%%%%%%%%%%%%%%%%%%
\begin{abstract} 
We used the asymmetric superconducting quantum interference device (SQUID) technique to extract the current phase relation (CPR)  of a Josephson junction with a 3D-topological insulator (TI) \ce{Bi2Se3} nanobelt as the barrier. The obtained CPR shows deviations from the standard sinusoidal CPR with a pronounced forward skewness. At temperatures below 200 mK, the junction skewness values are above the zero temperature limit for short diffusive junctions. Fitting of the extracted CPR shows that most of the supercurrent is carried by ballistic topological surface states (TSSs), with a small contribution of diffusive channels primarily due to the bulk. These findings are instrumental in engineering devices that can fully exploit the properties of the topologically protected surface states of 3D TIs.
\end{abstract}

\maketitle
%%%%%%%%%%%%%%%%%%%%%%% Introduction %%%%%%%%%%%%%%%%%%%%%%%%%%%%%
\vspace{0.2cm}
\noindent \textbf{1. Introduction\\}

Topological superconductivity and Majorana zero-energy modes  have attracted vast interest over the past few years owing to their potential for topologically-protected quantum information processing    \cite{Kitaev2002,Nayak2008,Sarma2015}. Hybrid devices involving a conventional s-wave superconductor (S) in proximity to a semiconducting nanowire with strong spin-orbit coupling   \cite{Lutchyn2010,Mourik2012} or an unconventional metal such as a 3D topological insulator (3D-TI)   \cite{Hasan2010,Fu2009,Tkachov2019} are expected to provide platforms for emulating and studying this exotic phenomena. One of the standard implementations of such hybrid devices includes S-3DTI-S junctions that exploit the topological surface states for hosting Majorana Bound States (MBSs). Over the past decade, such Josephson junctions based on TI materials have been fabricated and extensively studied experimentally   \cite{Williams2012,Cho2013,Galletti2014a,Kurter2015,Sochnikov2015,Wiedenmann2016a,Bocquillon2017,Li2018,Kunakova2019,LeCalvez2019,Ren2019,Schuffelgen2019,Kayyalha2019,Kayyalha2020,Kunakova2020,DeRonde2020,Stolyarov2020,Kim2020,Rosenbach2021,Bai2022,Schmitt2022,Fischer2022}. Here, Majorana physics manifests as peculiar properties of a part of the Andreev bound states (ABSs) carrying the Josephson current across the junction, namely Majorana bound states (MBSs). In an S-TI-S junction with multiple transport modes, MBSs are gapless even for not perfectly transparent S-TI interferences, and under proper conditions, they should show  a 4$\pi$ periodic current phase relation (CPR) coexisting  with a 2$\pi$ periodic CPR due to conventional ABS   \cite{Fu2009,Cook2012,Snelder2013,Tkachov2019}. The two periodicities  should be reflected in the total CPR of the junction, and by probing it one could get access to the unconventional physics of MBSs   \cite{Williams2012,Cho2013,Galletti2014a,Kurter2015,Sochnikov2015,Wiedenmann2016a,Bocquillon2017,Li2018,Kunakova2019,LeCalvez2019,Ren2019,Schuffelgen2019,Kayyalha2019, Kayyalha2020,Kunakova2020,DeRonde2020,Stolyarov2020,Rosenbach2021,Bai2022,Schmitt2022,Fischer2022}.
%%%%
%%%%
%%%%

The CPR of the junction can be probed using various DC and RF measurement techniques. These include current biased asymmetric DC-SQUIDs   \cite{DellaRocca2007,Kurter2015,Sochnikov2015,Spanton2017,Murani2017,Assouline2019,Kayyalha2020,Nichele2020,Mayer2020}, and magnetic field pattern measurements of single junctions   \cite{Williams2012,Ghatak2018,Chen2018,Assouline2019,Kononov2020}, phase-controlled junctions   \cite{Ren2019}, microwave-induced Shapiro steps   \cite{Kwon2004,Wiedenmann2016a,Dominguez2017,Bocquillon2017,Li2018,LeCalvez2019,Li2020,Park2021,Rosenbach2021,Fischer2022} and RF-SQUIDs coupled to microwave resonator readouts   \cite{Murani2019,Haller2022}. However, the critical point to note when looking for a $4 \pi$ periodic CPR of MBSs is that if the temporal variation of the phase across the junction is slower compared to the inelastic scattering time or the quasi-particle poisoning time, these processes will restore the $2\pi$ periodicity of the CPR   \cite{Fu2009,Badiane2011}. As a result, the recent studies aimed at detecting MBSs based on TI-junctions primarily focus on Shapiro step measurements at frequencies larger than any relaxation or poisoning rate   \cite{Wiedenmann2016a,Bocquillon2017,Li2018,LeCalvez2019,Rosenbach2021,Fischer2022} or microwave probing of phase-biased Josephson junctions   \cite{Murani2019}. The missing of odd integer Shapiro steps were reported, pointing toward the possible presence of $4\pi$ periodic modes in TI junctions   \cite{Wiedenmann2016a,Bocquillon2017,Li2018,LeCalvez2019,Rosenbach2021,Fischer2022,Kwon2004,Dominguez2017,Park2021}. 
%%%%
%%%%
%%%%

A significant obstacle in revealing MBSs using Josephson junctions based on 3D-TI like \ce{Bi2Se3}, \ce{Bi2Te3}, and \ce{Sb2Te3} is the coexistence of bulk states in addition to the topological surface states (TSSs), making the electrical transport analysis cumbersome  \cite{Hasan2010,Kong2011}. Compensation doping has been used to reduce the bulk contribution, however, at the expense of electron mobility   \cite{Cho2013,Schuffelgen2019,Kayyalha2019,Kayyalha2020,Stolyarov2020,DeRonde2020,Rosenbach2021,Bai2022,Schmitt2022}.
Another approach for reducing the bulk contribution to the electric transport is to increase the surface-to-volume ratio of the 3D TI  by growing the material in the shape of nanowires or nanobelts   \cite{Andzane2015,Kunakova2018,Kunakova2021}. Previous studies have shown high-quality interfaces between 3D-TI Bi$_2$Se$_3$ nanobelts and Al electrodes. These Josephson junctions show multiple Andreev reflections and large excess currents in the current-voltage characteristics   \cite{Kunakova2019,Kunakova2020}. Here, we further explore the properties of these junctions. Since the results we present here are based on DC measurements, we do not expect to observe any signature of MBS in the Josephson properties \cite{Fu2009,Badiane2011}. Rather, our study is aimed at characterizing ABS in 3D-TI nanobelt-based junctions.

%%%%  
%%%%
%%%%

In this work, we study the CPR of a \ce{Bi2Se3} nanobelt-based Josephson junction embedded in an asymmetric dc-SQUID. To keep the analysis simple, we focus on junctions in the short limit where the superconducting  coherence length $\xi$ is larger than the length of the junction $l$. Here the ABSs dispersion takes a simple form given by $E_n =\pm \Delta [1-\tau_n \sin^2(\varphi/2)]^{1/2}$, where $\Delta$ is the superconducting gap, $E_n$ and $\tau_n$ correspond to the Andreev level energy and transmission probability of the n$^{th}$ mode, respectively, and $\varphi$ is the phase difference across the junction   \cite{C.W.J.Beenakker1991}. The corresponding CPR of a short junction can be written as,
%%%%%%%%%%%%%%% Equation 1 %%%%%%%%%%%%%%%%%%
%%%%%%%%%%%%%%%%%%%%%%%%%%%%%%%%%%%%%%%%%%%%%%
\begin{equation}
\begin{split}
        I(\varphi) = & \frac{e\Delta(T)}{2\hbar}  \sum_{n=1}^{N} \frac{\tau_n \sin(\varphi) }{[1-\tau_n\sin^2(\varphi/2)]^{1/2}} \\
        & \times \tanh\left( \frac{\Delta (T)}{2k_B T} [1-\tau_n\sin^2(\varphi/2)]^{1/2}\right)  
\end{split}
\label{Eq1}
\end{equation}
%%%%%%%%%%%%%%%%%%%%%%%%%%%%%%%%%%
%%%%%%%%%%%%%%%%%%%%%%%%%%%%%%%%%%
where $T$ is the temperature, $\Delta(T)$ is the corresponding superconducting gap, $\hbar$, $k_B$, and $e$ are the reduced Planck constant, Boltzmann constant, and electron charge, respectively. In the above equation, the sum is taken over all transport modes in the TI junction. In TI nanobelts, spatial confinement along the transversal direction results in the formation of electronic sub-bands and a gap at the Dirac node     \cite{Kunakova2020,Rosenbach2021}. This prevents the observation of a perfectly transmissive transport mode. Even though we should not expect a mode with transparency 1, one should still be able to observe the contributions of transport modes with transparency close to one due to the peculiar linear Dirac dispersion in the surface states modes. The demonstration of these high transparency modes is the main objective of this paper.\\
%%%%
%%%%
%%%%

%%%%%%%%%%%%%%%%%%%%%%%%%%%%%%%%%%%%%%%%%%%%%%%%%%%%%%%%%%%%%%%%%%%%%%%%%%%%%%%%%%%%%
%%%%%%%%%%%%%%%%%%%%%%%%%%%%%%%% Fig 1 %%%%%%%%%%%%%%%%%%%%%%%%%%%%%%%%%%%%%%%%%%%%%%
%%%%%%%%%%%%%%%%%%%%%%%%%%%%%%%%%%%%%%%%%%%%%%%%%%%%%%%%%%%%%%%%%%%%%%%%%%%%%%%%%%%%%
 \begin{figure}
  \vspace{0.8cm}
\includegraphics[width=8.0cm]{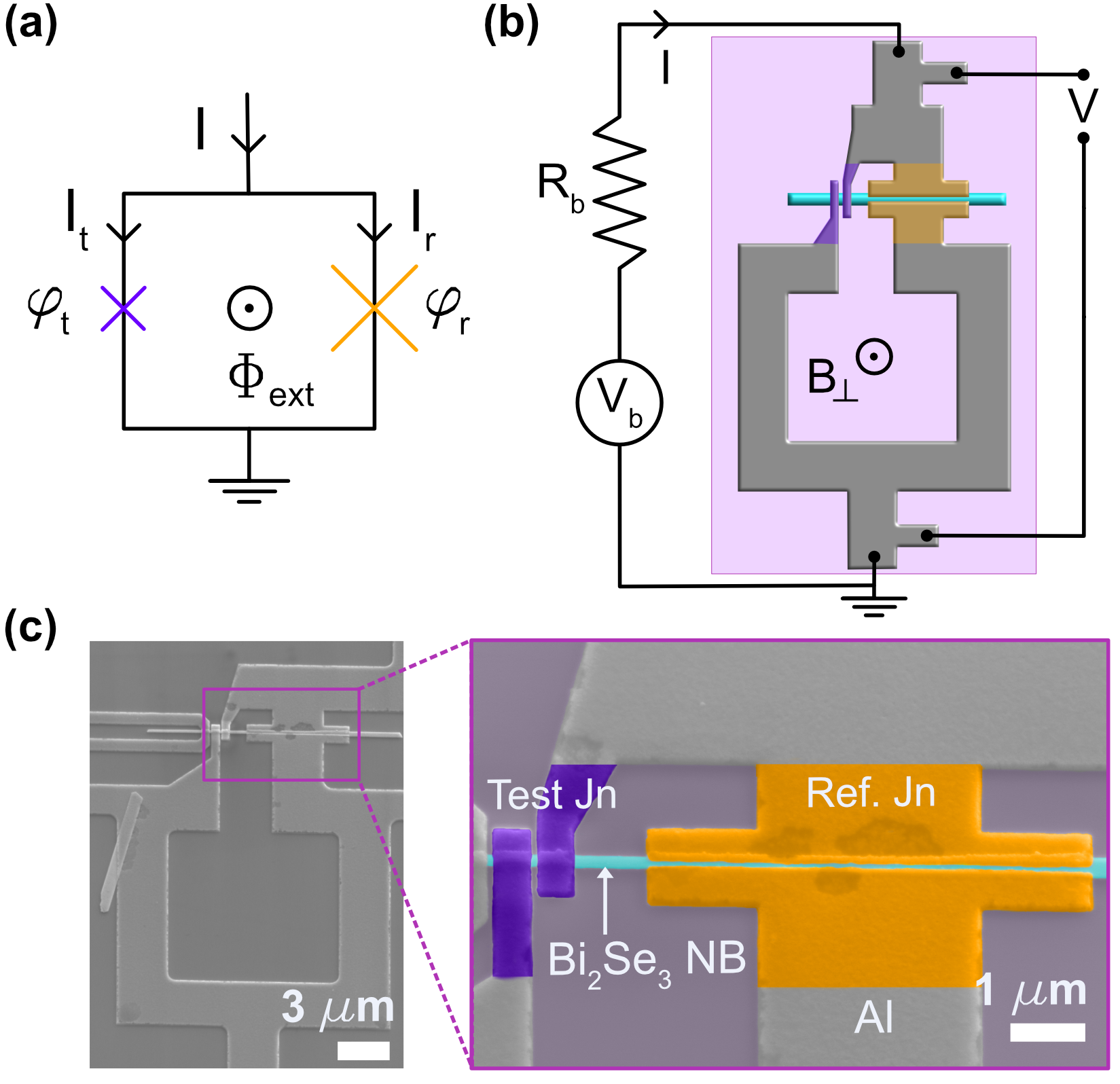}
\caption{(a) Sketch of the SQUID containing both the test (purple) and reference (orange) junction (b) Layout of the SQUID with both the test and reference junction made from the same \ce{Bi2Se3} nanobelt. The current-voltage characteristic is measured using a 4-point setup. (c) Scanning electron microscopy (SEM) image of the SQUID device (left panel). The right panel is a false-colored SEM image showing the test  (purple)  and reference(orange) junctions formed out of a \ce{Bi2Se3} nanobelt(cyan).}
\label{Fig1}
\end{figure}
%%%%%%%%%%%%%%%%%%%%%%%%%%%%%%%%%%%%%%%%%%%%%%%%%%%%%%%%%%%%%%%%%%%%%%%%%%%%%%%%%%%%%
%%%%%%%%%%%%%%%%%%%%%%%%%%%%%%%%%%%%%%%%%%%%%%%%%%%%%%%%%%%%%%%%%%%%%%%%%%%%%%%%%%%%%
%%%%%%%%%%%%%%%%%%%%%%%%%%%%%%%%%%%%%%%%%%%%%%%%%%%%%%%%%%%%%%%%%%%%%%%%%%%%%%%%%%%%%

%%%%%%%%%%%%%%%%%%%%%%% Methods %%%%%%%%%%%%%%%%%%%%%%%%%%%%%
\vspace{0.2cm}
\noindent \textbf{2. Methods}\\

The asymmetric dc-SQUID measurement is a powerful technique to extract the current phase relation (CPR) of Josephson junctions   \cite{DellaRocca2007,Nanda2017,Thompson2017,Kayyalha2019,Kayyalha2020}. Here, the  test junction with unknown CPR $I_t = I_{t,c}\cdot f(\varphi_t)$ is integrated into a dc-SQUID layout along with a reference junction with known CPR $I_r = I_{r,c}\cdot g(\varphi_r)$ and sufficiently higher critical current $I_{r,c}$, typically 15-20 times larger than the critical current of the test junction $I_{t,c}$, see  figure \ref {Fig1}.(a). Here $\varphi_t$ and $\varphi_r$ are the phases across the test and reference junctions, respectively. If the inductance of the SQUID loop, $L$,  is small enough such that the screening parameter  $\beta_L=\left(I_{r,c}+I_{t,c}\right)L/\Phi_0 \ll 1$, upon application of an external  magnetic flux, $\Phi_{ext}$, the phase across the test junction is given by $\varphi_t \simeq \varphi_r + 2\pi\Phi_{ext}/\Phi_0$, with  $\Phi_0$ the superconductive flux quantum   \cite{DellaRocca2007}. Due to the large asymmetry of the critical currents, the phase across the reference junction remains approximately constant in applied magnetic flux. Therefore the maximum critical current of the SQUID is obtained for $I_r = I_{r,c}$ at $\varphi_{r,max}$, where the current through the reference junction is maximized, while the phase across the test junction, $\varphi_t$, varies approximately linearly with $\Phi_{ext}$   \cite{DellaRocca2007}. Thus, the CPR of the test junction can be determined by subtracting the constant contribution of the reference junction, $I_{r,c}$, assuming a point-like junction, from the total  critical current of the  SQUID and going from flux to phase following the relation,
%%%%%%%%%%%%%%% Equation 2 %%%%%%%%%%%%%%%%%%
%%%%%%%%%%%%%%%%%%%%%%%%%%%%%%%%%%%%%%%%%%%%%%
\begin{equation}
\varphi_t \simeq \varphi_{r,max} + \frac{2\pi\Phi_{ext}}{\Phi_0}.
\label{Eq2}
\end{equation}
%%%%%%%%%%%%%%%%%%%%%%%%%%%%%%%%%%%%%%%%%%%%%%%
%%%%%%%%%%%%%%%%%%%%%%%%%%%%%%%%%%%%%%%%%%%%%%%
For a tunnel-like reference junction, one gets $\varphi_{r,max}=\pi/2$. However, for our devices, the reference junction is not a conventional tunnel junction. Indeed, nano-processing steps, especially the ones involving heating or etching, tend to alter the properties of TI materials. Therefore we opted to use only a single lithography step to reduce damage to the TI- junctions during device fabrication. In this case, even though the CPR of the reference junction is not known a priori, as long as $I_{r,c}\gg I_{t,c}$, one can still extract the CPR of the test junction. However, in this case $\varphi_{r,max}$ is not $\pi/2$. Earlier experiments have shown that this approach is reliable for extracting the CPR of the test junction   \cite{Nanda2017,Thompson2017,Mayer2020}.

We have realized Josephson junctions using \ce{Bi2Se3} nanobelts, grown by physical vapor deposition, which are at least 7-8 $\mu m$ long to be able to fabricate both the test and reference junction on the same nanobelt    \cite{Kunakova2019,Kim2020}. The fabrication process involves the dry transfer of nanobelts to a \ce{SiO2}/Si substrate followed by electron beam lithography (EBL) and metallization. Following our previous works, before e-beam evaporation of the Pt(3nm)/Al(80-100nm) electrodes, a mild  Ar ion milling is performed to remove the native oxide on the nanobelts  \cite{Galletti2014a,Kunakova2019,Kunakova2020}. SEM images of the SQUID device are shown in figure \ref{Fig1}.(c). Here, the length and width of the reference junction are defined by the separation between the two Al electrodes and the dimension of the electrodes along the longitudinal direction of the nanobelt, respectively. In contrast, the width of the test junctions is fixed by the width of the nanobelt (see figure \ref{Fig1}).\\
%%%%%
%%%%%
%%%%%
%%%%%%%%%%%%%%%%%%%%%%% Results and discussion %%%%%%%%%%%%%%%%%%%%%%%%%%%%%

\vspace{0.2cm}
\noindent \textbf{3. Results and discussion}\\

 We will focus on measurements from a single representative SQUID (BSH13 A3S2) formed from a \ce{Bi2Se3} nanobelt of width $w \simeq~$188 nm (from SEM in-lens image) and thickness $t \simeq~$48 nm (from AFM image, data not shown). Similar behavior has been observed in other devices (data not shown). The test junction has a length $l\simeq83~$nm, and the corresponding width and length of the reference junction are $\simeq5~\mu$m and $\simeq 80$~nm, respectively. For the ballistic case ($l<$ mean free path $\simeq$ 200 nm    \cite{Andzane2015}), the coherence length can be estimated using $\xi=\hbar v_F/\Delta'$, with $v_F\simeq 5\times 10^{5}~$m/s the Fermi velocity of the surface states in \ce{Bi2Se3},   \cite{Andzane2015,Kunakova2018} and $\Delta'$ the induced superconducting gap in the surface state. For a typical $\Delta'\simeq 135~\mu$eV extracted from single junction devices  \cite{Kunakova2019} we obtain $\xi\simeq 2.4~\mu$m, which is much longer than the length of our junctions, placing them in the short junction limit  \cite{C.W.J.Beenakker1991}. The SQUID loop line width is kept at 3~$\mu$m in most sections of the loop to minimize kinetic inductance contributions,  and thus have $\beta_L<1$. From the layout of the SQUID loop, by numerically solving the London-Maxwell equations we determined the effective area $A_{eff}\simeq 200~\mu$m$^2$. This results in a modulation period of  approximately  $10~\mu$T. From the same numerical calculations, we also extract a SQUID loop inductance value $L\simeq 29~$pH, of which  $\sim 27~$pH correspond to the geometric inductance, and the remaining $\sim 2~$pH are the kinetic contribution to loop inductance  \cite{Johansson2009}. Here we used a London penetration depth of $\lambda=70~$nm, typical for 100 nm thick Al films  \cite{Romijn1981}. Unless mentioned, all the measurements were carried out in a dilution refrigerator with a base temperature of 19~mK. The measurement lines are equipped with RC filters at the 4K stage and copper powder filters at the mixing chamber stage to minimize environmental noise/radiation reaching the device. 
 
%%%%
%%%%%%%%%%%%%%%%%%%%%%%%%%%%%%%%%%%%%%%%%%%%%%%%%%%%%%%%%%%%%%%%%%%%%%%%%%%%%%%%%%%%%
%%%%%%%%%%%%%%%%%%%%%%%%%%%%%%%% Fig 2 %%%%%%%%%%%%%%%%%%%%%%%%%%%%%%%%%%%%%%%%%%%%%%
%%%%%%%%%%%%%%%%%%%%%%%%%%%%%%%%%%%%%%%%%%%%%%%%%%%%%%%%%%%%%%%%%%%%%%%%%%%%%%%%%%%%%
 \begin{figure}
  \vspace{0.8cm}
\includegraphics[width=8.5cm]{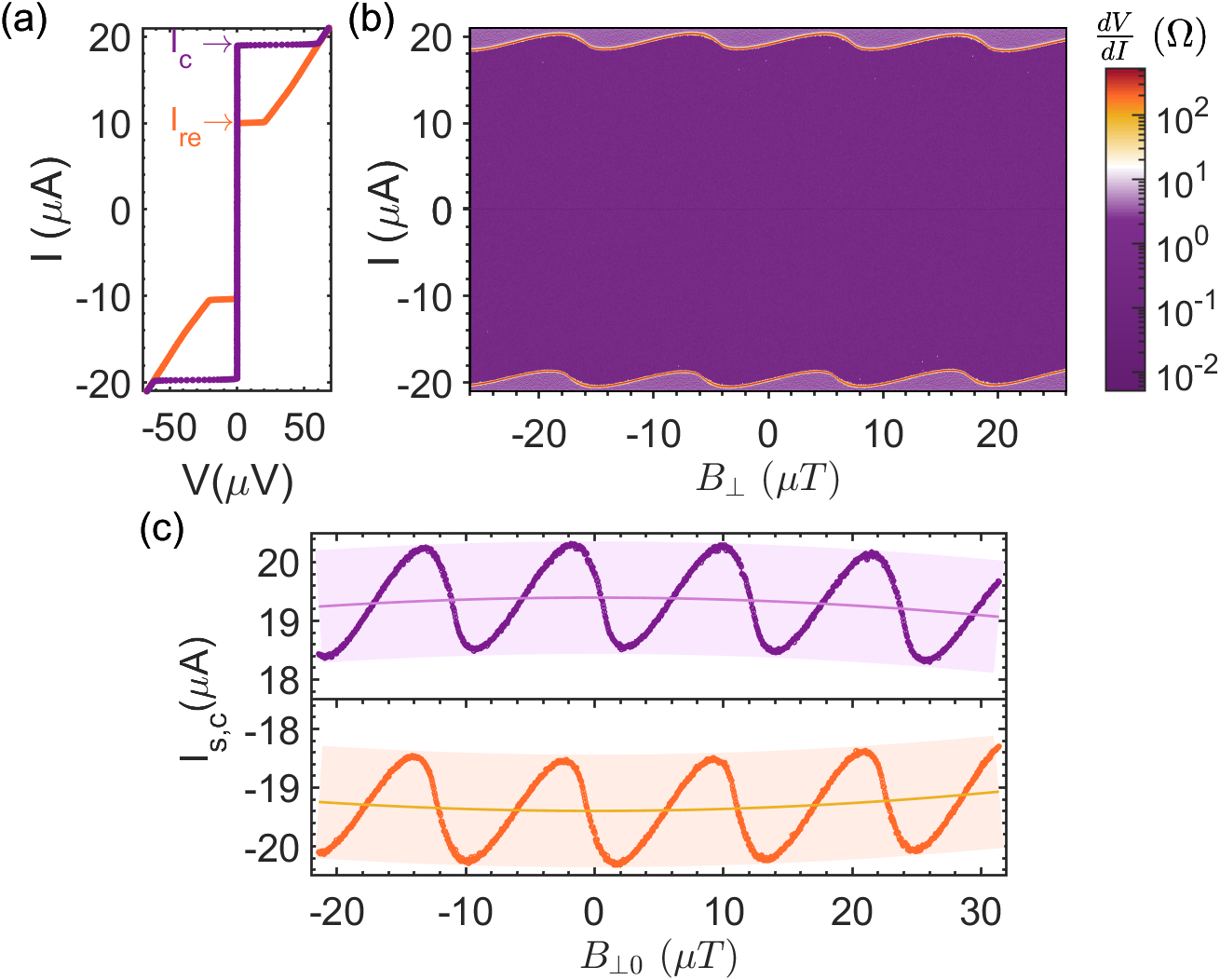}
\caption{ (a) Current-voltage characteristics at $T = 20~$mK with zero applied magnetic field. The arrows indicate the direction of the sweep (starting from zero), and the IVC shows hysteretic behavior . Here, $I_{c}$ and $I_{re}$ correspond to the critical current and retrapping current of the device, respectively. (b) Differential resistance of the SQUID as a function of bias current and externally applied magnetic field measured at $T = 20~$mK.  (c) The critical current of the SQUID for positive (upper panel) and negative (lower panel) bias current. The light-colored lines indicate the background envelope from the magnetic field pattern of the reference junction.  Here we adjusted the magnetic field data for a constant offset, $B_{off}$, to $B_{0\perp} = B_{\perp} + B_{off} $, with $B_{\perp}$ the applied magnetic field (see  panel (a)) and $B_{off}\simeq 5 \mu T $. Here, the offset was determined from the measured magnetic field position of the maxima of the background envelope (Fraunhofer pattern).
}
\label{Fig2}
\end{figure}
%%%%%%%%%%%%%%%%%%%%%%%%%%%%%%%%%%%%%%%%%%%%%%%%%%%%%%%%%%%%%%%%%%%%%%%%%%%%%%%%%%%%%
%%%%%%%%%%%%%%%%%%%%%%%%%%%%%%%%%%%%%%%%%%%%%%%%%%%%%%%%%%%%%%%%%%%%%%%%%%%%%%%%%%%%%
%%%%%%%%%%%%%%%%%%%%%%%%%%%%%%%%%%%%%%%%%%%%%%%%%%%%%%%%%%%%%%%%%%%%%%%%%%%%%%%%%%%%%
We measured the current-voltage characteristic (IVC) of the SQUID for various externally applied magnetic fields. The IVC of the TI SQUID at zero applied magnetic field is given in figure \ref{Fig2}.(a). Here, one can see the typical hysteretic IVC of Al-$\ce{Bi2Se3}$-Al junctions, and we attribute the origin of this hysteresis to heating effects  \cite{Courtois2008,Kunakova2019}. Now, to get the critical current($I_c$) of the device, we need to consider the bias sections of the IVC when the junction switches from the superconducting state to the resistive state (in both positive and negative bias directions), which are plotted in purple in  figure \ref{Fig2}.(a). For the rest of the analysis, we will ignore the sections of IVC where the junction goes back from the resistive state to the superconducting state (plotted in orange in figure \ref{Fig2}.(a)), as the switch occurs at the retrapping current ($I_{re}$), which is lower than the critical current of the device. Figure \ref{Fig2}.(b) shows the variation of the differential resistance $dV/dI$ of the TI-SQUID with respect to the applied bias current $I$ and external magnetic field $B_\perp$. Here, one can clearly see the modulations of the critical current from the asymmetric SQUID (bright lines).
%%%%
%%%%
%%%%

\begin{figure}
\centering
\includegraphics[width=8.5cm]{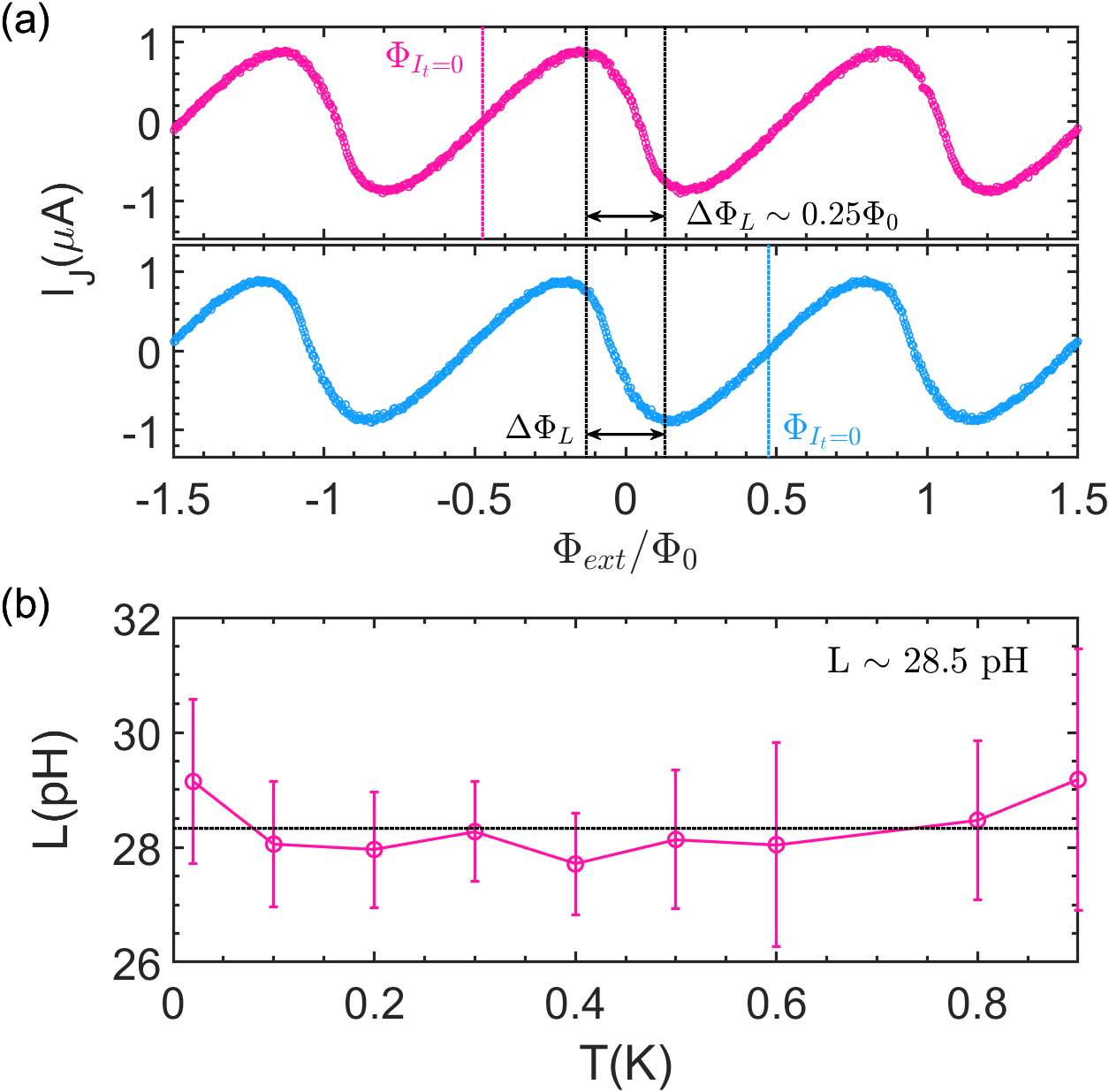}
\caption{(a) Extracted current-flux relation of the TI test junction at $T=20~$mK for both positive (magenta) and negative (blue) current bias directions of the asymmetric TI-SQUID and corresponding arrows indicate the location of maxima of C$\Phi$R. The shift in the location of positive and negative maxima from integer $\Phi_{ext}/\Phi_0$ positions due to finite inductance can be quantified in terms of $\Delta \Phi_L\ = L (I_{r,c} - I_{t,c})$, with $(I_{r,c} - I_{t,c})/2$ being the circulating current in the SQUID loop. (b) Loop inductance of the SQUID estimated using $\Delta \Phi_L$ as a function of bath temperature. The error bars represent the standard deviations of $L$ obtained from different pairs of the C$\Phi$R maxima corresponding to various integer $\Phi_{ext}/\Phi_0$ locations that are used for estimating $\Delta \Phi_L$. As one can see, L remains constant around the value of 29 pH confirming that it is dominated by the geometric inductance of the device.}
\label{Fig3}
\end{figure}

 Next, we determined the critical current of the SQUID, $I_{s,c}$, from the IVC for every applied magnetic field by setting a threshold voltage of $3~\mu$V as the criteria for detecting the switch from the superconducting state to the normal state. The resulting modulation of the positive and negative critical currents are shown as closed symbols in  figure \ref{Fig2}.(c). On top of the SQUID modulations, we observe a background envelope (solid lines) arising from the magnetic field modulation of the  critical current $I_{r,c}$ of the reference junction (Fraunhofer pattern).  Upon close examination, one could see that the maxima of the Fraunhofer patterns on both positive and negative sides occurred at the field of $-5 \mu T $. This means we have a constant offset in the magnetic field at the device. We see the same magnetic field offset in every measurement that we perform using the setup. To account for this shift, the magnetic field scale in figure \ref{Fig2}. (c) is offset to  $B_{0\perp} = B_{\perp} + B_{off} $. From the maxima of the  Fraunhofer pattern, we get $I_{r,c}$ to be 19.4 $\mu$A. Now, by removing the background due to the reference junction from the total response of the SQUID, one obtains the current modulations of the test junction ($I_{J}$) as a function of $B_{0\perp}$. By converting from magnetic field to flux, $\Phi_{ext}=B_{0\perp}A_{eff}$ using the observed modulation period of $\simeq 11.6\mu$T corresponding to one flux quantum $\Phi_0$, we obtain the current-flux relation (C$\Phi$R) of our test junction for the positive and negative bias currents, as shown in  figure \ref{Fig3 }. (a). From here, we get  $I_{t,c}\approx$ 880 nA, and the critical current asymmetry in our SQUID device is $I_{r,c}/I_{t,c}\approx$ 22, which is large enough for a proper CPR extraction. Using the simulated value of loop inductance, we can estimate the screening parameter $\beta_L\simeq 0.28$, which is not $\ll$ 1, and we will have to account for finite inductance effects when extracting the CPR.  
%%%%
%%%%%%%%%%%%%%%%%%%%%%%%%%%%%%%%%%%%%%%%%%%%%%%%%%%%%%%%%%%%%%%%%%%%%%%%%%%%%%%%%%%%%
%%%%%%%%%%%%%%%%%%%%%%%%%%%%%%%% Fig 3 %%%%%%%%%%%%%%%%%%%%%%%%%%%%%%%%%%%%%%%%%%%%%%
%%%%%%%%%%%%%%%%%%%%%%%%%%%%%%%%%%%%%%%%%%%%%%%%%%%%%%%%%%%%%%%%%%%%%%%%%%%%%%%%%%%%%
%%%%%%%%%%%%%%%%%%%%%%%%%%%%%%%%%%%%%%%%%%%%%%%%%%%%%%%%%%%%%%%%%%%%%%%%%%%%%%%%%%%%%
%%%%%%%%%%%%%%%%%%%%%%%%%%%%%%%%%%%%%%%%%%%%%%%%%%%%%%%%%%%%%%%%%%%%%%%%%%%%%%%%%%%%%
%%%%%%%%%%%%%%%%%%%%%%%%%%%%%%%%%%%%%%%%%%%%%%%%%%%%%%%%%%%%%%%%%%%%%%%%%%%%%%%%%%%%%
%%%%  
%%%%%%%%%%%%%%%%%%%%%%%%%%%%%%%%%%%%%%%%%%%%%%%%%%%%%%%%%%%%%%%%%%%%%%%%%%%%%%%%%%%%%
%%%%%%%%%%%%%%%%%%%%%%%%%%%%%%%% Fig 4 %%%%%%%%%%%%%%%%%%%%%%%%%%%%%%%%%%%%%%%%%%%%%%
%%%%%%%%%%%%%%%%%%%%%%%%%%%%%%%%%%%%%%%%%%%%%%%%%%%%%%%%%%%%%%%%%%%%%%%%%%%%%%%%%%%%%
\begin{figure*}
\centering
\includegraphics[width=16cm]{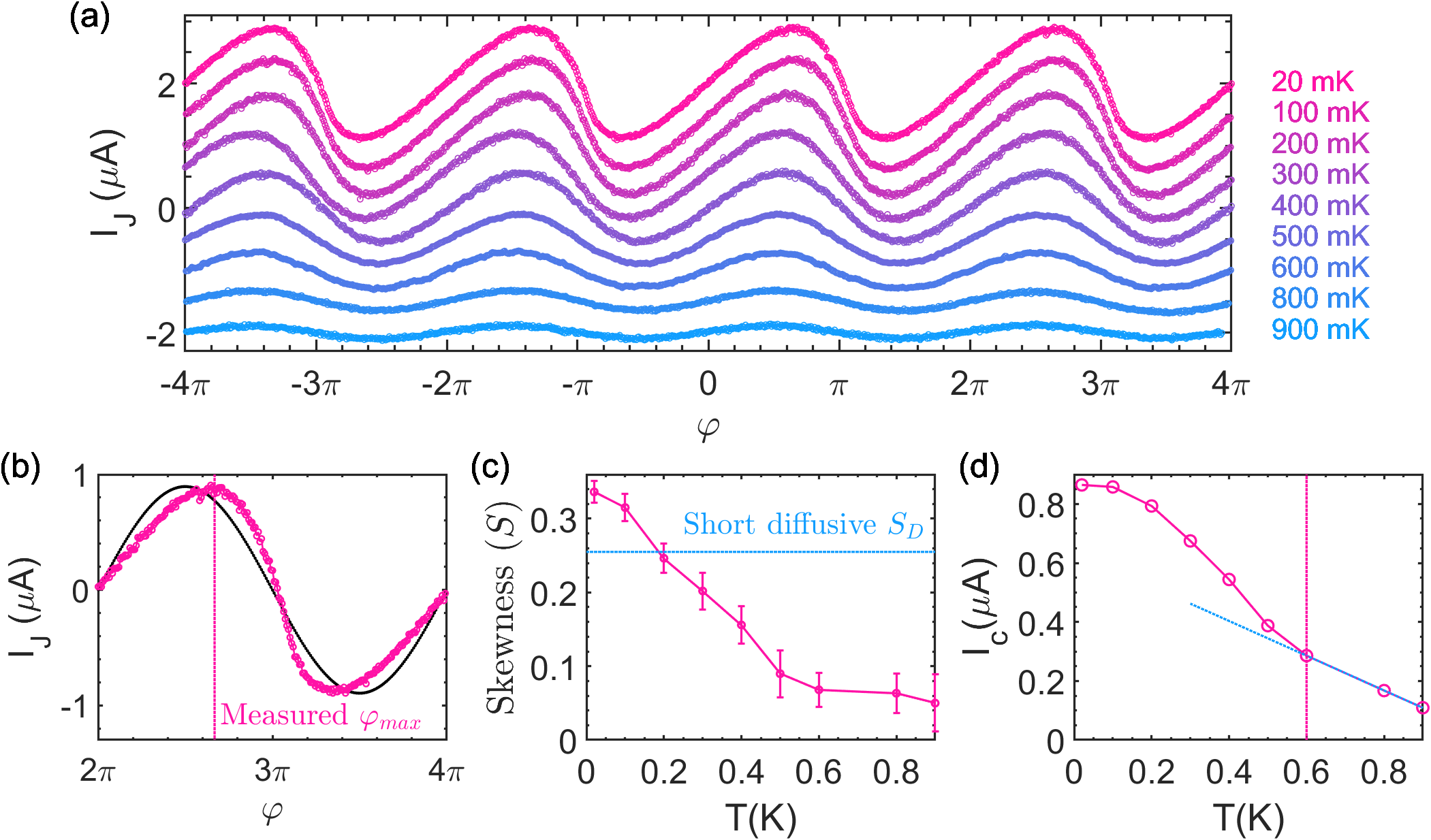}
\caption{(a) Evolution of the extracted CPR of the TI test junction with varying temperature. (b) Extracted CPR at 20 mK (magenta) along with a sinusoidal curve with the same amplitude (black) to emphasize the deviation of CPR from the conventional case. The vertical dotted magenta line indicates the location of $\varphi_{max}$ for the measured CPR. (c) Plot showing measured skewness of CPR vs. temperature. Below 200 mK the $S$ is greater than the zero temperature limit $S_D$, and the transport of the junction can not be explained fully by pure diffusive transport modes. (d) The critical current of the test junction as a function of temperature is shown. The open symbols are the measured data extracted from the CPRs shown in  panel (a). The dotted line  is the change in the slope of I$_c$ vs T curve, indicating the presence of two different types of transport modes with a higher (around  $T_c$ of Al,  1.15 K)  and a lower (around 600 mK) critical temperature.}
\label{Fig4}
\end{figure*}
%%%%%%%%%%%%%%%%%%%%%%%%%%%%%%%%%%%%%%%%%%%%%%%%%%%%%%%%%%%%%%%%%%%%%%%%%%%%%%%%%%%%%
%%%%%%%%%%%%%%%%%%%%%%%%%%%%%%%%%%%%%%%%%%%%%%%%%%%%%%%%%%%%%%%%%%%%%%%%%%%%%%%%%%%%%
%%%%%%%%%%%%%%%%%%%%%%%%%%%%%%%%%%%%%%%%%%%%%%%%%%%%%%%%%%%%%%%%%%%%%%%%%%%%%%%%%%%%%
%%%%

 In the following, we will look at the effect of inductance on the extracted C$\Phi$Rs and thereby estimate the inductance of the SQUID loop experimentally. From equation \ref{Eq2}, one can see that for the negligible inductance case, when the external flux is an integer multiple of $\Phi_0$, we get $\varphi_t \simeq \varphi_{r,max}$. Now, if we assume the CPRs of the test and reference junction have the same functional shape, $f(\varphi)\simeq g(\varphi)$, the C$\Phi$Rs will have the maxima at external flux values which are an integer multiple of $\Phi_0$. But as seen from figure\ref{Fig3}.(a),  the maxima (in both positive  and negative current directions as indicated by the magenta and blue arrows respectively) are offset from integer $\Phi_{ext}/\Phi_0$ values. This shift in the C$\Phi$Rs is due to the field produced by the circulating current in the SQUID loop, and can be accounted for by modifying the equation for the phase across the test junction as $\varphi_t = \varphi_r + 2\pi\Phi/\Phi_0$ with $\Phi = \Phi_{ext}-L\left[I_r(\varphi_r)-I_t(\varphi_t)\right]/2$ the total magnetic flux through the SQUID loop  \cite{Nichele2020}, assuming the inductance is distributed equally ($L/2$) among the two arms of the SQUID. This shift depends on the direction of the bias current through the SQUID loop, and in our case, the C$\Phi$Rs are shifted to the left(right) for positive(negative) bias current. Now to quantify this, for every integer $\Phi_{ext}/\Phi_0$ positions, one can define a parameter $\Delta\Phi_L$ as the distance between the observed location of the maxima pair corresponding to positive and negative current bias (see the black dotted lines in figure \ref{Fig3}.(a)). From the equation for $\varphi_t$ in the finite inductance case, it is straightforward to see that $\Delta\Phi_{L} = L (I_{r,c} - I_{t,c})$, with $(I_{r,c} - I_{t,c})/2$ the circulating current in the loop when the SQUID critical current is maximized  \cite{Nichele2020}. In our device,  by considering four pairs of peaks in the C$\Phi$R (only three are shown for clarity), we obtain $\Delta \Phi_L\simeq 0.28~\Phi_0$ corresponding to a loop inductance value of roughly 29~pH. This value is in good agreement with the $L$ value that we find from our numerical simulations \cite{Johansson2009}. In  figure \ref{Fig3}.(b) we show the extracted SQUID loop inductance for various temperatures up to 900~mK. Here, within the error bars of our data, we do not see a considerable increase in the inductance of the SQUID loop with temperature, confirming that the geometric inductance dominates the loop inductance in the entire temperature range.
%%%%%%
%%%%%%
%%%%%%

To convert the magnetic flux to the phase drop across the test junction, one should, in principle use equation \ref{Eq2}. However,  as discussed above, this equation is only valid for zero loop inductance. To account for the finite inductance, we need to subtract from equation \ref{Eq2} the flux shift due to the circulating current. This can be obtained by subtracting the magnetic flux value at which the current of the test junction goes to zero $\varphi_{t}=2\pi(\Phi_{ext}-\Phi_{I_t=0})/\Phi_0$ flux (see  magenta and blue dotted lines in figure \ref{Fig3}.(a)), which accounts for both $\varphi_{r,max}$ and the contribution of the circulating current. The resulting CPRs at various temperatures are shown in  figure \ref{Fig4}.(a). Note that these are not the true CPRs of the test junction. This is because for finite $\beta_L$, the flux to phase conversion is not completely linear as in equation \ref{Eq2}, and deviations from the linear trend occur around odd multiple integers of $\pi$ (see discussion below)  \cite{Nichele2020}. Also, from here on, we use $\varphi$ instead of $\varphi_t$ to indicate the phase across the test junction extracted assuming linear flux to phase conversion  and reserve $\varphi_t$ to indicate the actual phase across the test junction.
%%%%%%
%%%%%%
%%%%%%

As seen in   figure \ref{Fig4}.(a), at low temperatures, the extracted CPRs are forward skewed and evolve into a more sinusoidal CPR at higher temperatures. One commonly used method to quantify the departure of CPRs from a conventional sinusoidal CPR  is by defining the skewness, $S = (2\varphi_{max}/\pi)-1$, where $\varphi_{max}$ corresponds to the phase at which the critical current of the junction is reached  \cite{Nanda2017,Kayyalha2020,Haller2022}. The maximum skewness $S=1$ with $\varphi_{max}=\pi$ is achieved for transmission probability $\tau_n=1$ (see equation \ref{Eq1}), whereas for $\tau_n\ll 1$ one obtains $S=0$ (sinusoidal CPR). We find that at 20 mK,  $\varphi_{max}$ = 0.67$\pi$, which corresponds to a skewness $S=0.34$ (see  figure \ref{Fig4}.(b)). Here, we can rule out any inductance contributions to the position of $\varphi_{max}$, since  in the limit of a small screening parameter the CPR is affected only in a small phase region around odd  integer multiples of $\pi$ (see below). In  figure \ref{Fig4}.(c) we show the monotonic decrease of the skewness $S$ with increasing temperature, asymptotically approaching zero for higher temperatures. We can now compare the skewness parameter to the predictions for different junction regimes. For a short diffusive junction, $\xi<l$, with $\xi=\sqrt{\hbar D/\Delta'}$ the coherence length, $D$ the diffusion constant, and $\Delta'$ the induced superconducting gap, one expects skewness $S_D=0.255$ (see dotted line in  figure \ref{Fig4}. (c))  \cite{Haller2022}. We clearly observe that for temperatures below 200 mK, the transport in our TI junction can not be described by pure diffusive transport. In the limit of Josephson transport coming only from the surface TI channels (TI surface modes) one can estimate the skewness by calculating the CPR using equation \ref{Eq1} and the transmission probabilities $\tau$  following   \cite{Titov2006}, assuming an extreme chemical potential mismatch between the TI channel and the  TI covered by the Al electrode,

%%%%%%%%%%%%%%% Equation 3 %%%%%%%%%%%%%%%%%%
%%%%%%%%%%%%%%%%%%%%%%%%%%%%%%%%%%%%%%%%%%%%%%
\begin{equation}
\begin{split}
        \tau_n = & \frac{k_n^2}{k_n^2\cos^2(k_nl)+k_F^2\sin^2(k_nl)}  
\end{split}
\label{Eq3}
\end{equation}
%%%%%%%%%%%%%%%%%%%%%%%%%%%%%%%%%%%%%%%%%%%%%%%
%%%%%%%%%%%%%%%%%%%%%%%%%%%%%%%%%%%%%%%%%%%%%%%
with quantized electron momentum $k_n = \sqrt{k_F^2-k_y^2}$ along longitudinal direction, where $k_F$ is the Fermi wave vector and $k_y$ the quantized transverse momentum given by $k_y = 2\pi(n+\frac{1}{2}) /C$ with $n = 1,2,3..$ and C being the circumference of the nanobelt ($2(w+t)$). Here we obtain $S=0.41$ at 20~mK. The fact that the observed skewness parameter is smaller than the one expected for the Josephson current carried exclusively by the surface state suggests that bulk states contribute to the overall current as well. This is supported by the temperature dependence of the critical current (I$_c$) of the test junction shown in  figure \ref{Fig4}.(d). The open symbols are the measured data extracted from the CPRs shown in  figure \ref{Fig4}.(a). As indicated by the dashed lines, around 600 mK we observe a drastic change in the slope of the I$_c$ vs $T$ curve that might be due to the presence of two different types of Josephson transport channels with different $T_C$ values as reported in   \cite{Schuffelgen2019,Rosenbach2021}. The Josephson current contribution with higher $T_C$ (same as the Al electrode, $T_C= 1.15~$K) can be attributed to the ballistic TSSs, while the Josephson current contribution that vanishes fast around 600 mK could be attributed to the diffusive transport modes due to bulk states. This is further supported by the low bulk mobility values we extract from magnetotransport data  \cite{Kunakova2018}.
 
%%%%
%%%%%%%%%%%%%%%%%%%%%%%%%%%%%%%%%%%%%%%%%%%%%%%%%%%%%%%%%%%%%%%%%%%%%%%%%%%%%%%%%%%%%
%%%%%%%%%%%%%%%%%%%%%%%%%%%%%%%% Fig 5 %%%%%%%%%%%%%%%%%%%%%%%%%%%%%%%%%%%%%%%%%%%%%%
%%%%%%%%%%%%%%%%%%%%%%%%%%%%%%%%%%%%%%%%%%%%%%%%%%%%%%%%%%%%%%%%%%%%%%%%%%%%%%%%%%%%%
\begin{figure*}
\includegraphics[width=17cm]{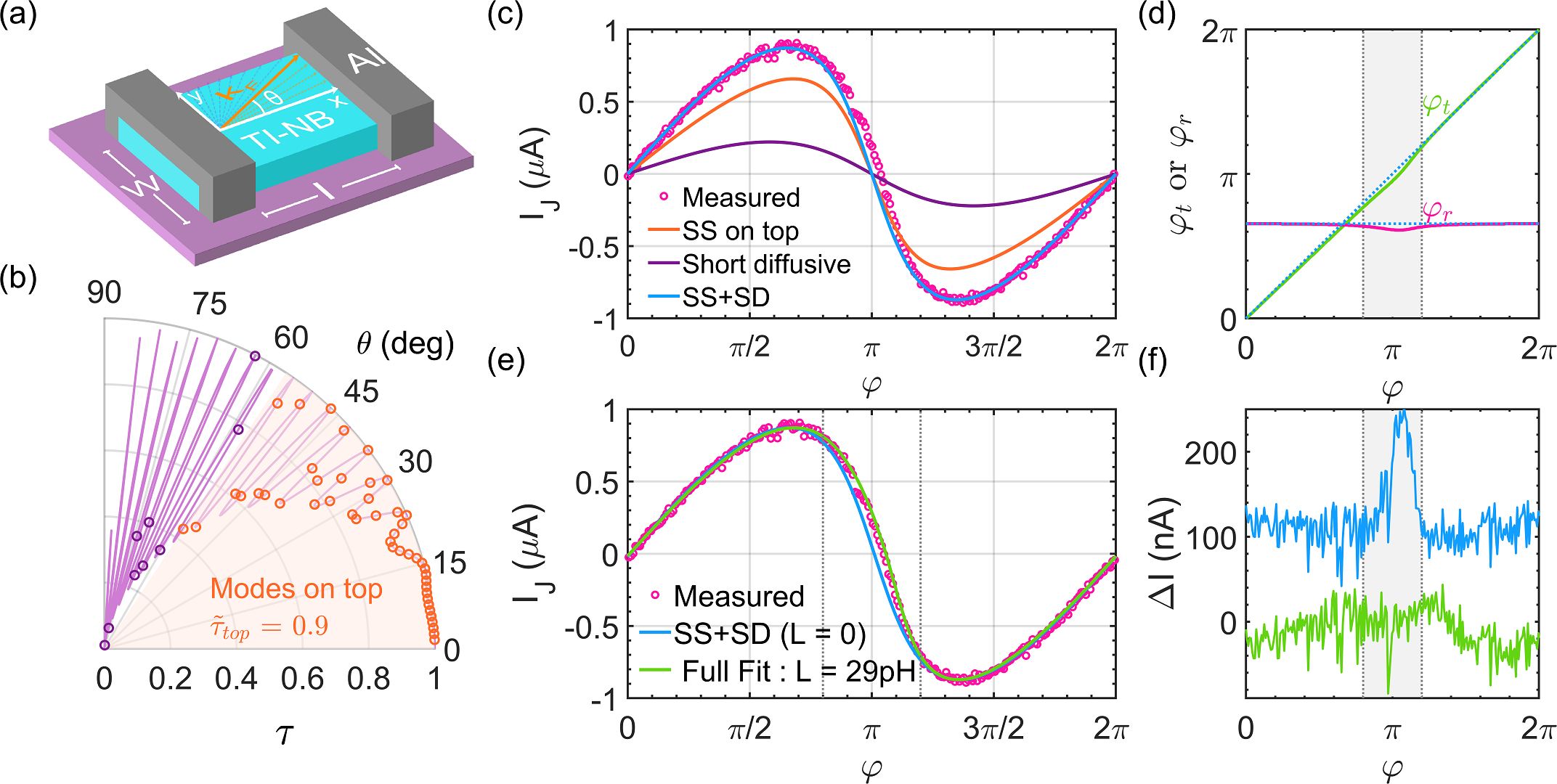}
\caption{(a) Sketch of an S-TI-S  junction showing quasi-particle trajectories in terms of the angle $\theta$ made with longitudinal momentum axis with $\theta = \tan^{-1}{(\frac{k_x}{k_y})}$. (b) Angle-dependent transmission probabilities (Fabry-P\'erot like resonance s) of various transport modes in our junction, assuming extreme  chemical potential mismatch at the interface. The orange circles correspond to the modes which are fully contained on the top surface  of the junction, and the purple dots correspond to the modes that propagate around the nanobelt's perimeter. (c) Measured CPR of the junction at 20 mK (magenta circles) along with fitted curves   assuming zero inductance, $L= 0$. The orange curve is the current contribution from  the ballistic TSSs on the top surface of the junction, and  the purple line shows the diffusive contributions to  the CPR. The sum of both ballistic and diffusive modes is given as the blue curve. (d) Variation of $\varphi_t$ and $\varphi_r$ with respect to $\varphi$, the extracted phase across the junction assuming a linear variation of phase with $\Phi_{ext}$. Most of the deviation occurs around $\varphi\simeq\pi$ (see the shaded region, $0.8\pi-1.2\pi$). (e) Measured CPR of the junction (magenta circles) along with  the fitted curve  considering a finite loop inductance of $L= 29~$pH (green line)~. The zero inductance fit is given for comparison (blue).  The region in between the dashed gray lines corresponds to regions mostly affected by inductance as in panel (d), whereas $\varphi_{max}$ of the CPR lies outside this region. (f)  Difference or residual between the measured data and fit ($\Delta I = I_J-I_{fit}$). The data are offset for clarity. For the zero inductance case, there is a peak in $\Delta I$ around $\pi$, indicating the  deviation of the fitted curve around $\pi$ from the measurement data. The green curve corresponds to  the fit including the loop inductance resulting in a pronounced reduction of the peak around $\pi$. The shaded region is the same as in panel (d).}
\label{Fig5}
\end{figure*}
%%%%%%%%%%%%%%%%%%%%%%%%%%%%%%%%%%%%%%%%%%%%%%%%%%%%%%%%%%%%%%%%%%%%%%%%%%%%%%%%%%%%%
%%%%%%%%%%%%%%%%%%%%%%%%%%%%%%%%%%%%%%%%%%%%%%%%%%%%%%%%%%%%%%%%%%%%%%%%%%%%%%%%%%%%%
%%%%%%%%%%%%%%%%%%%%%%%%%%%%%%%%%%%%%%%%%%%%%%%%%%%%%%%%%%%%%%%%%%%%%%%%%%%%%%%%%%%%%
%%%%

%%%%
%%%%
%%%%
To estimate the individual contributions from the bulk states and surface states to the total Josephson current, we fit  the CPR measured at 20 mK (see closed symbols in  figure \ref{Fig5}.(c)) with a two-band model. For the Josephson current carried by the TSSs we again consider quantized transport modes with transmission probabilities dictated by the geometry of the device, following   \cite{Titov2006} (see equitation \ref{Eq3}). 
After removing an oxide thickness of 5 nm from the width and thickness, we find for the nanoribbon used in the SQUID, C = 452 nm. Following our previous work, we only consider the modes that travel on the top surface of the TI-belt to be ballistic, as the modes that go around the circumference of the nanobelt suffer from poor mobilities due to the interface with the substrate  \cite{Kunakova2018,Kunakova2020,Kunakova2021} and/or the paths are longer than the phase coherence length. For the junction geometry, assuming a $k_F$ of 0.55 nm$^{-1}$ based on magnetotransport measurements performed by us on similar \ce{Bi2Se3} nanobelts \cite{Andzane2015,Kunakova2018,Kunakova2021}, we find that in total there are 39 modes arising from TSSs. Out of these, 29 of them travel on the top surface and should be contributing to the ballistic portion of the supercurrent. In   figure \ref{Fig5}. (b) we show the polar plot of the transmission probabilities as a function of (continuous) angle $\theta$ (see solid line) using equation \ref{Eq3}, displaying  Fabry-P\'erot-like resonance features  in the form of lobes with transmission probabilities very close 1 at certain $\theta$ values related to the geometry of the device  \cite{Li2018,Rosenbach2021}. The transmission probabilities of the individual quantized transport modes are shown as open symbols.   The orange open symbols correspond to the $\tau$ values of modes that are on the top surface of the junction (up to $\theta\approx $ 48$ ^\circ$). More than half of these modes have $\tau$ values close to 1, resulting in an average value of 0.92 for the modes on the top surface. The modes with $\theta$ above 48$^\circ$ (purple open symbols) that go around the circumference are not ballistic as discussed in   \cite{Kunakova2020}. So we describe the Josephson current carried by modes that go around the nanobelt and those carried by the bulk states with a diffusive multi-mode model using the Dorokhov distribution of transmission probabilities   \cite{Kos2013} given by,
%%%%%%%%%%%%%%% Equation 3 %%%%%%%%%%%%%%%%%%
%%%%%%%%%%%%%%%%%%%%%%%%%%%%%%%%%%%%%%%%%%%%%%
\begin{equation}
   \rho(\tau) = \frac{\pi\hbar G_N}{2e^2} \frac{1}{\tau\sqrt{1-\tau}}
\end{equation}
where $G_N$ is the normal-state conductance for diffusive modes. Since the value of $G_N$ is unknown, we take it as a fitting parameter in our analysis. Now, the CPR contribution from the diffusive modes is calculated by replacing the sum in equation \ref{Eq1} with an integral in the interval $\tau = [0,1]$  covering the full range of transmission probabilities. 
%%%%%%%%%%%%%%%%%%%%%%%%%%%%%%%%%%%%%%%%%%%%%%%
%%%%%%%%%%%%%%%%%%%%%%%%%%%%%%%%%%%%%%%%%%%%%%%
%%%%%
%%%%%
%%%%%

The fitted contributions to the CPR of our junction at 20 mK are shown in figure \ref{Fig5}.(c). To fit the CPR, we had to assume a  temperature (T$_{fit}\simeq 195$~mK), which is higher than the bath temperature of 20mK. This discrepancy may have its origins in elevated quasi-particle temperatures or additional current noise  typically observed in SNS junctions   \cite{DellaRocca2007,Akazaki2005}. The details about noise contributions are beyond the scope of this work and will be described elsewhere. The best fit is obtained when out of the total critical current of $\approx$ 880 nA (see magenta circles corresponding to measured data or blue line corresponding to the sum of ballistic and diffusive parts) of the test junctions, $\approx$ 657 nA are carried by ballistic TSSs on the top of the junction (orange line), and  $\approx$ 223 nA are carried by the diffusive transport modes (purple line). The presence of diffusive transport modes with lower skewness will reduce the skewness of the   overall CPR   of the junction as compared to transport carried entirely by the TSS, in agreement with our experiment.
%%%%%
%%%%%
%%%%%

Finally, we discuss the influence of the finite inductance on the extracted CPR. As one can clearly see in  figure  \ref{Fig5}.(c), the measured CPR does not cross zero at $\varphi_{t}=\pi$. This is due to the breakdown of the linear mapping between external flux and phase across the test junction around phase values of odd integer multiples of $\pi$. To resolve this and obtain better flux to phase conversion, one must solve the equation $\varphi_t = \varphi_r + 2\pi\Phi/\Phi_0$ for each value of $\Phi_{ext}$, including the finite SQUID loop inductance, to get the pairs of phase values $\varphi_r$ and $\varphi_t$, that maximize the current through the SQUID loop  \cite{Nichele2020}. The variation of $\varphi_r$ and  $\varphi_t$, calculated for a finite inductance value of $L=29~$pH, with respect to $\varphi$ is given in figure \ref{Fig5}.(d). Here one can see that the curves show deviations around $\pi$ (shaded region) from the expected linear behavior (dashed blue lines) corresponding to zero inductance case. A similar deviation visible in the difference between the measured CPR and the fitted curve is given in figure \ref{Fig5}.(f) in the form of a peak around $\pi$. Now using $\varphi_t$ values, that include inductance effects, we can reproduce the measured CPR relation more accurately and  reproduce the zero crossing of the CPR at a phase value slightly larger than $\pi$   as seen in figure \ref{Fig5}.(e). This is also reflected in the residual from the fit including the finite inductance of the SQUID loop (see figure \ref{Fig5}.(f)). In fact, the peak around $\pi$ disappears, leaving behind mostly the noise from the measurement. Finally, the true CPR of our TI-junction is represented by the blue curve corresponding to the zero inductance fit  in figure \ref{Fig5}.(c) and (e), and the deviation of the measured CPR from the theorectically expected CPR is simply caused by not fully satisfying  the condition $\beta_L\ll1$. Here, we note that the finite inductance value does not significantly affect the phase position of the maximum of the CPR (see  figure \ref{Fig5}.(e)), since the position of the maximum and mininmum of the CPR are outside the phase region (in between the dashed gray lines in Fig. \ref{Fig5} (d),(e), and(f)) where $\varphi_t$ deviates from the linear dependence on $\varphi$. Therefore, the various skewness values we extracted earlier are still valid, indicative of the short quasi-ballistic nature of our junction.

%%%%%%
%%%%%%
%%%%%%

%%%%%%%%%%%%%%%%%%%%%%% Conclusion %%%%%%%%%%%%%%%%%%%%%%%%%%%%%
\vspace{0.2cm}
\noindent \textbf{4. Conclusions}\\

To conclude, we extracted the CPR of Al-\ce{Bi2Se3}-Al junctions formed out of 3D-TI nanobelts using asymmetric SQUID measurements.   We observe a skewed CPR due to the TSSs  hosting transport modes with high transmission probabilities. We found that our junctions are, in the short, quasi-ballistic regime, with most supercurrent being carried by ballistic TSSs. However, to fit the extracted CPR, one has to consider both ballistic and diffusive contributions. Therefore, reducing the number of transport modes in these junctions is essential, especially the diffusive bulk contributions, in order to ensure fewer ambiguities in future experiments aimed at  detecting MBSs using 3D-TI materials-based devices.\\

%%%%
%%%%
%%%%%
\vspace{0.5cm}
\noindent\textbf{Data availability statement\\}

\noindent The main data that support the findings of this study are included
within the article. Additional data are available from the corresponding author upon reasonable request.\\

\vspace{0.2cm}
\noindent\textbf{Acknowledgement\\}

\noindent This work was supported by the European Union's H2020 under the Marie Curie Actions (No. 766025-QuESTech). This work has been supported by the European Union’s Horizon 2020 Research and Innovation Program (Grant Agreement No. 766714/HiTIMe).  K. Niherysh acknowledges the financial support of the “Strengthening of the capacity of doctoral studies at the University of Latvia within the framework of the new doctoral model”, identification no. 8.2.2.0/20/I/006.\\

\vspace{0.2cm}
\noindent\textbf{ORCID iDs}\\

\noindent A. P. Surendran \orcidlink{0000-0002-0949-4145} \href{https://orcid.org/0000-0002-0949-4145}{https://orcid.org/0000-0002-0949-4145}
\\
\noindent D. Montemurro \orcidlink{0000-0001-8944-0640} \href{https://orcid.org/0000-0001-8944-0640}{https://orcid.org/0000-0001-8944-0640}\\
\noindent G. Kunakova \orcidlink{0000-0003-0243-2678} \href{https://orcid.org/0000-0003-0243-2678}{https://orcid.org/0000-0003-0243-2678}\\
\noindent X. Palermo \orcidlink{0000-0001-9997-3053} \href{https://orcid.org/0000-0001-9997-3053}{https://orcid.org/0000-0001-9997-3053}\\
\noindent K. Niherysh \orcidlink{0000-0002-9861-9957 }\href{https://orcid.org/0000-0002-9861-9957}{https://orcid.org/0000-0002-9861-9957}\\
\noindent E. Trabaldo \orcidlink{0000-0002-0188-6814} \href{https://orcid.org/0000-0002-0188-6814}{https://orcid.org/0000-0002-0188-6814}\\
\noindent D. S. Golubev \orcidlink{0000-0002-0609-8921} \href{https://orcid.org/0000-0002-0609-8921}{https://orcid.org/0000-0002-0609-8921}\\
 \noindent J. Andzane \orcidlink{0000-0002-9802-6895} \href{https://orcid.org/0000-0002-9802-6895}{https://orcid.org/0000-0002-9802-6895}\\
 \noindent D. Erts \orcidlink{0000-0003-0345-8845} \href{https://orcid.org/0000-0003-0345-8845}{https://orcid.org/0000-0003-0345-8845}\\
\noindent F. Lombardi \orcidlink{0000-0002-3478-3766} \href{https://orcid.org/0000-0002-3478-3766}{https://orcid.org/0000-0002-3478-3766}\\
 \noindent T. Bauch \orcidlink{0000-0002-8918-4293} \href{https://orcid.org/0000-0002-8918-4293}{https://orcid.org/0000-0002-8918-4293}
 
\bibliography{Ref_library}
\end{document}